\documentclass[aps,prl,twocolumn,superscriptaddress,showpacs,amsmath,amssymb,preprintnumbers]{revtex4}

\usepackage{graphicx,amssymb}

\begin{document}


\title{Quantum Phase Interference and N\'eel-Vector Tunneling in Antiferromagnetic Molecular Wheels}

\author{O. Waldmann}
\email[E-mail: ]{oliver.waldmann@physik.uni-freiburg.de}

\affiliation{Physikalisches Institut, Universit\"at Freiburg, D-79104 Freiburg, Germany}

\author{T. C. Stamatatos}
\affiliation{Department of Chemistry, University of Florida, Gainesville, FL 32611-7200, USA}

\author{G. Christou}
\affiliation{Department of Chemistry, University of Florida, Gainesville, FL 32611-7200, USA}

\author{H. U. G\"{u}del}
\affiliation{Department of Chemistry and Biochemistry, University of Bern, 3012 Bern, Switzerland}

\author{I. Sheikin}
\affiliation{Grenoble High Magnetic Field Laboratory, CNRS, BP 166, 38042 Grenoble Cedex 9, France}

\author{H. Mutka}
\affiliation{Institut Laue-Langevin, 6 Rue Jules Horowitz, BP 156, F-38042 Grenoble Cedex 9, France}

\date{\today}

\begin{abstract}
The antiferromagnetic molecular wheel Fe$_{18}$ of eighteen exchange-coupled Fe$^\texttt{III}$ ions has been studied by
measurements of the magnetic torque, the magnetization, and the inelastic neutron scattering spectra. The combined data
show that the low-temperature magnetism of Fe$_{18}$ is very accurately described by the N\'eel-vector tunneling (NVT)
scenario, as unfolded by semiclassical theory. In addition, the magnetic torque as a function of applied field exhibits
oscillations that reflect the oscillations in the NVT tunnel splitting with field due to quantum phase interference.
\end{abstract}


\maketitle

Although magnetism is a priori quantum mechanical, observation of genuine quantum phenomena such as tunneling and phase
interference in magnets is difficult. This is because in ferro- or ferrimagnets, the low-temperature magnetism and
dynamics are by and large well described by the magnetization vector $\textbf{M}$ obeying classical equations, although
for magnetic molecules (e.g. Mn$_{12}$, Fe$_8$), tunneling of the magnetization and phase interference have been
observed \cite{Mn12_Fe8}. Antiferromagnets, however, exhibit zero magnetization in the ground state, and observation of
quantum behavior is even more challenging.

Antiferromagnets can be described by the N\'eel vector $\textbf{n} = (\textbf{M}_A - \textbf{M}_B)/(2M_0)$, with
sublattice magnetizations $\textbf{M}_A$, $\textbf{M}_B$ of length $M_0$ (Fig.~1b). In three dimensions, they exhibit
long-range N\'eel order, but domains enable thermally activated or even quantum fluctuations of the N\'eel vector
\cite{Shp07}. Nanosized, single-domain antiferromagnetic (AFM) clusters would provide clean systems to study
N\'eel-vector dynamics, but mono-dispersity is then a key issue. In small particles with weak magnetic anisotropy, the
N\'eel vector can rotate \cite{And84,OWCCR}, while with strong anisotropy it is localized in distinguishable
directions, but fluctuates by quantum tunneling, i.e., N\'eel-vector tunneling (NVT) \cite{Gun95,Bar90,Chi98}.
Typically, there are two tunneling paths (Fig.~1c) and interference occurs due to the phase of the wave function. This
gives rise to characteristic oscillations in the tunnel splitting as function of applied magnetic field, which would be
observable in static measurements such as magnetization \cite{Chi98}.

Initial attempts to establish NVT with the ferritin protein \cite{Gid95} unfortunately proved controversial owing to
polydisperse samples and the presence of uncompensated magnetization \cite{ferritin}. Recent attempts with the AFM
molecular wheels CsFe$_8$ and Fe$_{10}$ were encouraging steps forward \cite{NVT,San05}, but were criticized with
arguments that the tunneling actions $S_0/\hbar$ are too small and the NVT picture only approximately valid
\cite{San05}. The latter attempts were stimulated by the theoretical prediction of NVT in such wheels \cite{Chi98}, and
that its observation would be assisted by the monodispersity and crystallinity inherent in molecular compounds, and
their resulting well defined structural and magnetic parameters \cite{Taf94,Gat94,OWCCR,Kin06}.

\begin{figure}[b]
\includegraphics[scale=0.93]{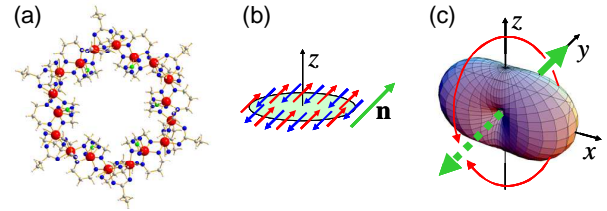}
\caption{\label{fig:one} (color online)
(a) Molecular structure of Fe$_{18}$ (red: Fe$^{III}$). (b) Schematics of the classical ground-state configuration of
the Fe spins (red, blue arrows) with the N\'eel vector indicated (green arrow, length not to scale). (c) Shape of the
potential $V(\textbf{n})$ in the high-field regime ($g \mu_B B / \hbar \omega_0 = 1.2$). The two tunneling paths from
$\textbf{n} = +y$ to $-y$ are indicated.}
\end{figure}

We here present measurements of the magnetic torque (${\bf \tau} = \textbf{M} \times \textbf{B}$), magnetization, and
inelastic neutron scattering (INS) on the AFM molecular wheel
[Fe$_{18}$(pd)$_{12}$-(pdH)$_{12}$(O$_2$CEt)$_6$(NO$_3$)$_6$](NO$_3$)$_6$, or Fe$_{18}$ (Fig.~1a). The combined data
show that the low-temperature magnetism in Fe$_{18}$ is very accurately described by the NVT scenario, as unfolded by
semiclassical theory \cite{Chi98}. Moreover, the torque is found to exhibit wiggles that directly reflect the
oscillations in the tunnel splitting with field, or quantum phase interference indeed.

The generic spin Hamiltonian for AFM wheels is
\begin{equation}
 \hat{H} = -J \sum^{N}_{i=1}{ \hat{\textbf{S}}_i \cdot \hat{\textbf{S}}_{i+1} }
  + D \sum^{N}_{i=1}{ \hat{S}^2_{iz} }
  + g \mu_B \hat{\textbf{S}} \cdot \textbf{B},
\end{equation}
with nearest-neighbor Heisenberg interactions of strength $J < 0$ (periodic boundaries assumed), a uniaxial single-ion
magnetic anisotropy of strength $D$ along the wheel axis $z$, and a Zeeman interaction ($N = 18$ is the number of ions,
$\hat{\textbf{S}}_i$ is the spin operator of the $i$th ion with spin $s = 5/2$ for Fe$^\texttt{III}$). The
path-integral treatment - or semiclassical theory - of Eq.~(1) has provided analytic results for the low-$T$ magnetism
and describes its dynamics in terms of the intuitive picture of NVT \cite{Chi98}.

For $D > 0$ and a field $B$ along $x$ perpendicular to $z$, the N\'eel-vector dynamics are governed by the potential
$V(\textbf{n}) = N/|8J|(g \mu_B B )^2 n_x^2 + Ns^2D n_z^2$, with hard-axis anisotropies due to the field $||x$ and the
magnetic anisotropy $||z$. Thus, $V(\textbf{n})$ has minima at $\textbf{n} = \pm y$ separated by an energy barrier
(Fig.~1c). NVT between these two classical states occurs if the energy barrier is much larger than the attempt
frequency $\hbar \omega_0 = s \sqrt{8D|J|}$, producing a tunnel splitting $\Delta$ in the quantum-mechanical energy
spectrum. In the high-field regime $g \mu_B B > \hbar\omega_0$ the anisotropy barrier is smaller than that due to the
field. Hence, the N\'eel vector tunnels via the $z$ axis, through the anisotropy barrier $U = Ns^2D$, but the tunneling
from e.g. $+y$ to $-y$ may be either along the 'upper' or 'lower' path in Fig.~1c. However, the phase of the wave
function acquires a term $\pi/2$ due to quantum fluctuations plus a topological term $\pi N g \mu_B B/|4J|$
proportional to the field. Thus, with increasing field one repeatedly tunes through destructive and constructive
interference, and the tunnel splitting oscillates between zero and a maximum according to
\begin{equation}
  \Delta(B) = \Delta_0 \left|\sin \left( \pi \frac{N g \mu_B B}{4|J|}  \right)   \right|,
\end{equation}
with $\Delta_0 = 8 \hbar\omega_0 \sqrt{S_0/h} \exp(-S_0/\hbar)$ and the tunneling action $S_0/\hbar = Ns\sqrt{2D/|J|}$.
$S_0/\hbar$ is a crucial parameter: the above condition for NVT to occur translates into $S_0/\hbar
> 2$ (henceforth we specifiy $D/|J|$ in terms of $S_0/\hbar$). Importantly, the tunneling affects also
the ground-state energy $\varepsilon_0(B) = \varepsilon(B) - \Delta(B)/2$, where $\varepsilon(B)$ varies smoothly,
typically quadratically with $B$ \cite{Chi98}. Hence, the oscillations in $\Delta(B)$ can be detected, at zero
temperature, via the static magnetization or torque \cite{Chi98}.

Fe$_{18}$ was synthesized as in Ref.~\cite{Kin06}. Torque and magnetization data were recorded on single crystals with
excellent quality. The shapes were three-sided pyramidal, allowing accurate alignment of the crystals. Torque was
measured with a CuBe cantilever inserted into the M10 magnet at the Grenoble High Magnetic Field Laboratory equipped
with an Oxford dilution fridge. Magnetic field was in the $xz$ plane, with an angle $\theta$ between field and $z$.
Magnetization was measured with the same device, but displaced from the magnet's field center, which resulted in a
Faraday force proportional to the magnetization. Despite very careful alignment, a small torque contribution to the
signal could not be avoided; hence only the parts of the data are shown where the estimated torque contribution is
smaller than 5\%. A smooth background is also present, which is estimated to be smaller than 5\%. INS spectra of a
polycrystalline sample of undeuterated Fe$_{18}$ were recorded on the spectrometer IN5 at the Institut Laue-Langevin,
Grenoble. Data were corrected for detector efficiency by a vanadium standard. Initial wavelength was $\lambda =
5.0$~{\AA} and temperatures were $T$ = 1.5 and 4.2~K. Spectra were summed over all detector banks. Experimental
resolution at the elastic line was 95~$\mu$eV.

The magnetization and torque curves vs field measured at $T = 0.1$~K are shown in Figs.~2a and 2c for fields (nearly)
parallel to $z$. Both curves exhibit staircase-like steps at regular field intervals, as is common in AFM wheels
\cite{Taf94,Cor99,OWCCR}. However, for fields perpendicular to $z$, the magnetization, after a first broadened step,
increases linearly with field, with only weak features apparent (Fig.~2b). The torque is even more striking: after a
first step it exhibits wiggles at higher fields (Fig.~3a) - this is unprecedented for molecular magnetic clusters in
general, and AFM wheels in particular \cite{wiggles}. The wiggles in the torque directly correspond to oscillations in
the NVT splitting (\emph{vide infra}), and their observation is the main result of this work.

\begin{figure}
\includegraphics[scale=1]{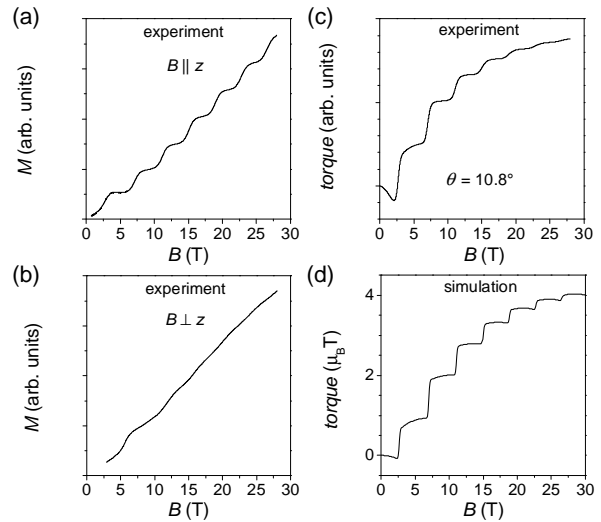}
\caption{\label{fig:two}
Magnetization vs field for (a) $B \| z$ and (b) $B \perp z$, and torque vs field for $B$ nearly parallel to
$z$ as (c) measured and (d) simulated using Eq.~(3) ($T = 0.1$~K).}
\end{figure}

The sign of the torque (Figs.~2c, 3a) determines $D > 0$. As pointed out in Ref.~\cite{Chi98}, the semiclassical theory
then predicts a dichotomy  in the magnetization as a key feature of NVT, i.e., a conventional staircase-like behavior
for parallel fields but a linear curve with weak features for perpendicular fields - exactly as observed in Fe$_{18}$.
This is a first, strong indication of the accuracy of the NVT picture in Fe$_{18}$. The field derivative of the
parallel magnetization (Fig.~4a) and the positions of the maxima (steps in the magnetization) shown in Fig.~4b
demonstrate the regular field intervals of the steps. The dependence of the first torque step on the field angle,
$B_1(\theta)$, is presented in Fig.~4c (features at higher fields show similar dependencies, but with much weaker
variation). In AFM wheels studied before, the angle dependence is well described by $B_1(\theta) = a + b(\cos^2\theta -
1/3)$, which was also observed for the wheels CsFe$_8$ and Fe$_{10}$ with the largest $S_0/\hbar$ reported to date
\cite{NVT,Car05}. Fe$_{18}$ shows a more pronounced angle dependence indicating a substantially larger $S_0/\hbar$.
Figure~4d presents the INS spectra at 1.5 and 4.2~K. Four transitions I, II, iii, and iv at ca. 0.3, 1.0, 0.8, and
1.35~meV, and a weak feature v at ca. 0.6~meV are observed for positive energy transfer. At negative energy transfer,
the expected anti-Stokes lines I' and iv' are visible. From the temperature dependence, peaks I and II are cold
transitions, and iii, iv, and v are hot transitions.

The combined data establish an AFM wheel, and Eq.~(1) with a large $S_0/\hbar$. The molecular structure of Fe$_{18}$
means that not all exchange paths are identical, as assumed in Eq.~(1). However, the energies of the low-lying states
relevant in our experiments are not affected, to first order, by variations in the exchange constants \cite{EPL}. $J$
should thus be taken as the average, as is also the case for the anisotropy constant $D$. Thus, with a modulation in
the exchange and anisotropy parameters, Eq.~(1) is the appropriate model to describe the low-$T$ behavior of Fe$_{18}$
(and therewith also the models discussed next).

\begin{figure}
\includegraphics[scale=1]{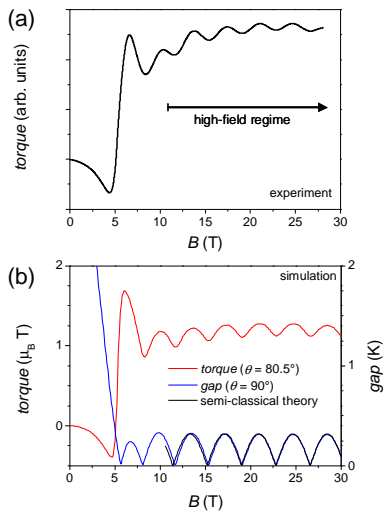}
\caption{\label{fig:three} (color online) Torque vs field for $B$ nearly
perpendicular to $z$ as (a) measured and (b) simulated ($\theta = 80.5^\circ$, $T = 0.1$~K). Panel (b) also
shows the gap between the two lowest quantum states vs $B$ as simulated with Eq.~(3) (blue curve) and
predicted by semiclassics, Eq.~(2) (black curve).}
\end{figure}

Since the Hilbert space for Fe$_{18}$ is huge ($\approx 10^{14}$), preventing direct use of Eq.~(1), our quantitative
analysis employs two approximations, which capture the low-energy i.e. low-$T$ physics well. First, the AFM
sub-lattices are replaced by spins $\hat{\textbf{S}}_A = \sum_{i \in A} \hat{\textbf{S}}_i$ and $\hat{\textbf{S}}_B =
\sum_{i \in B} \hat{\textbf{S}}_i$, each of length $Ns/2 = 45/2$. This gives
\begin{equation}
 \hat{H}_{AB} = -j \hat{\textbf{S}}_A \cdot \hat{\textbf{S}}_B
  + d \left( \hat{S}^2_{Az} +  \hat{S}^2_{Bz} \right)
  + g \mu_B \hat{\textbf{S}} \cdot \textbf{B},
\end{equation}
with $j = a_1 J$ and $d = b_1 D$ \cite{EPL}. The Hilbert space of Eq.~(3) is small (2116), which permits exact
numerical diagonalization. Second, we will use the analytical results of the semiclassical treatment of Eq.~(1)
\cite{Chi98}.

Diagonalizing Eq.~(3) and fitting to the data of Figs.~{4b-4d} gave $j = -5.1(1)$~K, $d = 0.021(1)$~K ($g = 2.0$). This
parameter set reproduces all our data with high accuracy ($\pm$ few percent), i.e., the fields of the magnetization
steps (Fig.~4b), the angle dependence $B_1(\theta)$ (Fig.~4c), the INS spectra (Fig.~4d), and the torque curves for
nearly parallel (Figs.~2c, 2d) and perpendicular fields (Fig.~3). Hence, Eq.~(3), and therewith Eq.~(1), describes the
low-energy sector of Fe$_{18}$ very accurately. The observed features in the torque (and magnetization) are less sharp
than in the simulations (Fig.~3), and possible reasons for this include (i) additional, very weak terms in the spin
Hamiltonian, e.g., Dzyaloshinski-Moriya interactions, which would lead to a rounding of the gap function and hence the
torque, and (ii) a small distribution (5\%) in the $j$ and $d$ values ($j$ and $d$ strain), which would result in a
distribution of step positions and hence broadening.

\begin{figure}
\includegraphics[scale=1]{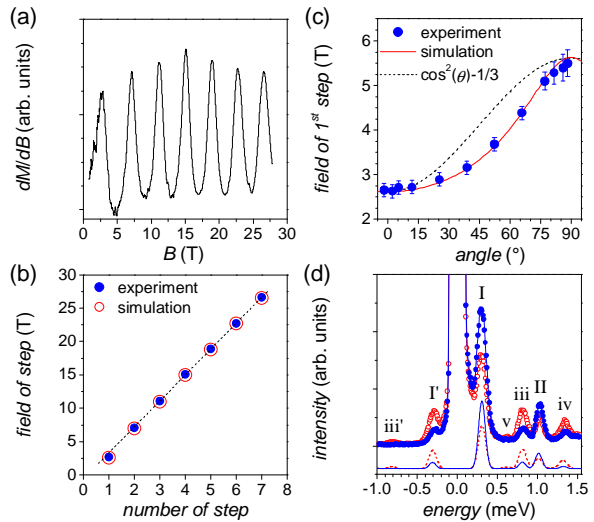}
\caption{\label{fig:four} (color online)
(a) $dM/dB$ vs field for $B \| z$ ($T = 0.1$~K). (b) Fields of the maxima in $dM/dB$ as measured
(full circles) and simulated (open circles). (c) Field of the first torque step vs angle, $B_1(\theta)$, as
observed (circles), expected in the strong-exchange limit (dashed line), and simulated with Eq.~(3) (red line). (d) INS
spectra at 1.9 (full blue circles) and 4.2~K (open red circles). Blue solid and red dashed lines show the simulated
spectra (elastic line not included in simulation).}
\end{figure}

Determining $J$ and $D$ now requires knowing $a_1$, $b_1$. In principle, $a_1$ and $b_1$ should be determined such that
the exact low-energy spectrum of Eqs.~(1) and (3) match \cite{EPL}. This is not trivial for Fe$_{18}$. Quantum
Monte-Carlo techniques yielded $a_1 = 0.2721$ \cite{Eng06}, and extrapolating $b_1$ for $N = 6, 8, 10$ to Fe$_{18}$
\cite{EPL} yields the estimate $b_1 \approx 0.07$. Hence, $J = -19$~K and $D \approx 0.3$~K, which are typical values
for ferric wheels. With these one gets $S_0/\hbar = 8.0$.

We now turn to semiclassical theory. First a subtle point needs to be addressed. The above method of determining $a_1$,
$b_1$ is certainly correct, but we observe that semiclassical theory yields quantitatively accurate results only if one
uses the $a_1$, $b_1$ values predicted by itself. This is expected from self-consistency considerations. Hence,
semiclassical formulae should be used with "corrected" $J$, $D$ values as deduced with the semiclassical values
$a^{sc}_1 = 2/9$, $b^{sc}_1 = 0.1065$ \cite{class}. The tunneling action then becomes $S_0/\hbar = 5.90$, which is the
value to be used \cite{S0}.

Henceforth we consider the high-field regime $g \mu_B B > \hbar\omega_0$, or $B > 10.6$~T for Fe$_{18}$. The criterion
for NVT is $S_0/\hbar > 2$, which is well fulfilled by $S_0/\hbar = 5.90$. The expectation value of the N\'eel vector
along $\pm y$ is $\langle0|n_y|0\rangle^2 \approx 1 - (S_0/\hbar)^{-1}$ \cite{Chi98}, hence in Fe$_{18}$ the N\'eel
vector is localized to 83\%. Another figure of merit is the ratio $\Delta_0/U$ of maximal tunnel splitting vs barrier
height. For Fe$_{18}$, $\Delta_0 = 0.320$~K and $U = 22.2$~K, thus $\Delta_0/U = 0.014$; the tunnel splitting is
exponentially small as expected for a tunneling scenario. Figure~3b displays the energy gap between the two lowest
levels for perpendicular fields as calculated with Eq.~(3) and the semiclassical formula Eq.~(2). The excellent
quantitative agreement finally demonstrates the high accuracy of the semiclassical NVT theory for Fe$_{18}$.

With reliable values for $j$, $d$, $S_0/\hbar$, and the accuracy of our modeling of Fe$_{18}$ established, we now
discuss the wiggles in the torque for nearly perpendicular fields. Figure~3b shows the calculated field dependencies
for the energy gap and the torque. In the high-field regime, the energy gap (= tunnel splitting) exhibits the typical
sinus-half-wave oscillation due to quantum interference \cite{Chi98}. Comparison with the torque is striking: in the
high-field regime the torque clearly follows the tunnel splitting, except for a smooth offset. The torque has not yet
been calculated by semiclassical theory, but the result for $M_x$ \cite{Chi98} inspires a refinement, which we checked
numerically to describe the torque. For $\theta \approx 90^\circ$, we write the magnetization as $\textbf{M} = g \mu_B
B (F_x,0,\cos\theta F_z)$ and the torque as $\tau = g \mu_B B \cos\theta(F_z - F_x)$, where, at 0~K,
\begin{equation}
  F_{\frac{x}{z}} = \frac{B}{B_0} - \frac{1}{2}
  + \frac{\pi N \Delta_0}{8|J|} (-1)^n \cos{\left( \pi \frac{B}{B_0} \pm \phi_{\frac{x}{z}} \right)} \mp
  f_{\frac{x}{z}}.
\end{equation}
Here we used $B_0 = 4|J|/(N g \mu_B)$ and a factor $(-1)^n$ such that $|\sin()|=(-1)^n \sin()$ ($n$ numbers
magnetization/torque steps). As compared to the semiclassical formula for $M_x$ we have, to grasp weak but important
effects of magnetic anisotropy, introduced shifts $\phi_x(B)$, $\phi_z(B)$ to take into account that the steps do not
occur exactly at fields $n B_0$, and functions $f_x(B)$, $f_z(B)$ to account for a smooth offset. Numerically we find
that $\phi_x$, $\phi_z$, $f_x$, and $f_z$ vary with field roughly as $B^{-1}$, in accord with expectation, such that
after some rearrangements
\begin{equation}
  \tau =  \frac{\pi N}{4|J|} \Delta_0 \left|\sin \left( \pi \frac{B}{B_0} \right) \right| g(B) \cos \theta  + f(B),
\end{equation}
where $g(B)$ and $f(B)$ are smooth, essentially constant functions of $B$. The torque provides only indirect evidence
for the size of the NVT splitting, but Eq.~(5) proves that it does directly probe the oscillations in the NVT
splitting, as demonstrated before in Fig.~3b. At non-zero temperatures, the sine term in Eq.~(5) should be multiplied
by $\tanh(\Delta/2k_BT)$, which has the effect of smearing out sharp features, or of rounding the torque wiggles. This
is why the simulated torque curve in Fig.~3b is less sharp at the zeros than the gap.

In summary, the combined magnetic torque, magnetization, and INS data, and their analyzes by two models, demonstrate
that the low-temperature magnetism in Fe$_{18}$ is in accord with the NVT picture of semiclassical theory. Most
importantly, the observed features in the magnetic torque have been demonstrated to directly reflect oscillations in
the NVT splitting, i.e., quantum phase interference. An interesting point is the number of electrons involved in the
tunneling. In Fe$_{18}$, the total spin on each AFM sub-lattice has a length of 45/2, i.e., 90 electrons are involved.
Hence, the disconnectivity, which is a measure for the "macroscopiness" of a quantum effect \cite{Leg02}, is 90. This
makes Fe$_{18}$ one of the most mesoscopic molecular systems exhibiting magnetic quantum tunneling.

\begin{acknowledgments}
Financial support by the Deutsche Forschungsgemeinschaft (DFG) and the USA National Science Foundation (CHE-0414555)
are acknowledged.
\end{acknowledgments}


\begin{references}

\bibitem{Mn12_Fe8}
 D. Gatteschi, R. Sessoli, J. Villain, Molecular Nanomagnets (Oxford University Press, Oxford, U.K., 2006);
 J. R. Friedman \emph{et al.}, Phys. Rev. Lett. \textbf{76}, 3830 (1996);
 W. Wernsdorfer and R. Sessoli, Science \textbf{284}, 133 (1999).

\bibitem{Shp07}
  O. G. Shpyrko \emph{et al.}, Nature \textbf{447}, 68 (2007).

\bibitem{And84}
  P. W. Anderson, Basic Notions of Condensed Matter Physics (Benjamin/Cummings Publishing Co., 1984).

\bibitem{OWCCR}
  O. Waldmann, Chem. Coordin. Rev. \textbf{249}, 2550 (2005).

\bibitem{Gun95}
  E. M. Chudnovsky and  J. Tejada, Macroscopic Quantum Tunneling of the Magnetic Moment (Cambridge University Press, 1998).

\bibitem{Bar90}
 B. Barbara and E. Chudnovsky, Phys. Lett. A \textbf{145}, 205 (1990).

\bibitem{Chi98}
 A. Chiolero and D. Loss, Phys. Rev. Lett. \textbf{80}, 169 (1998).

\bibitem{Gid95}
 S. Gider \emph{et al.} Science \textbf{268}, 77 (1995); J. Tejada \emph{et al.}, Phys. Rev. Lett. \textbf{79}, 1754 (1997);

\bibitem{ferritin}
 J. Tejada, Science 272, 424 (1996); A. Garg, \emph{ibid}, p. 424; S. Gider \emph{et al.}, \emph{ibid} p. 425.

\bibitem{NVT}
  O. Waldmann, C. Dobe, H. Mutka, A. Furrer, H. U.  G\"udel, Phys. Rev. Lett. \textbf{95}, 057202 (2005).

\bibitem{San05}
 P. Santini \emph{et al.}, Phys. Rev. B \textbf{71}, 184405 (2005).

\bibitem{Taf94}
 K. L. Taft \emph{et al.}, J. Am. Chem. Soc. \textbf{116}, 823 (1994).

\bibitem{Gat94}
 D. Gatteschi, A. Caneschi, L. Pardi, R. Sessoli, Science \textbf{265}, 1054 (1994).

\bibitem{Kin06}
 P. King, T. C. Stamatatos, K. A. Abboud, G. Christou, Angew. Chem. Int. Ed. \textbf{45}, 1 (2006).

\bibitem{Cor99}
 A. Cornia \emph{et al.}, Angew. Chem. Int. Ed. \textbf{38}, 2264 (1999).

\bibitem{wiggles}
 "Wiggles" in the torque vs field were observed before in Mn-[3$\times$3], Cr$_7$Ni, and
Cr$_7$Zn, but these are not due to NVT but a direct, first-order mixing of $|S,-S\rangle$ and $|S+1,-S-1\rangle$ levels
by the magnetic anisotropy \cite{QMO,Car05}. Experimentally, the situation in these molecules is unambiguously
distinguished from the observations in Fe$_{18}$ by the fact that the wiggles occur for both nearly parallel and
perpendicular fields, in striking contrast to Fe$_{18}$, where ordinary staircase-like profiles are observed for
parallel fields. The situation in Fe$_{18}$ hence cannot be confused with that found in the above molecules. It is
noted that in high fields NVT cannot be described by a mixing of $|S,-S\rangle$ and $|S+1,-S-1\rangle$ states.

\bibitem{QMO}
 O. Waldmann \emph{et al.}, Phys. Rev. Lett. \textbf{92}, 096403 (2004).

\bibitem{Car05}
  S. Carretta \emph{et al.}, Phys. Rev. B \textbf{72}, 060403(R) (2005).

\bibitem{EPL}
  O. Waldmann, Europhys. Lett. \textbf{60}, 302 (2002).


\bibitem{Eng06}
 L. Engelhardt and M. Luban, Phys. Rev. B \textbf{73}, 054430 (2006).

\bibitem{class}
  $a^{sc}_1 = 4/N$, $b^{sc}_1 = 2Ns^2/[Ns(Ns+2)-3]$ \cite{Chi98,EPL}.

\bibitem{S0} Unfortunately, this point was not considered before, and previously reported $S_0/\hbar$ values
    should be used with care.

\bibitem{Leg02}
 A. J. Leggett, Phys.: Condens. Matter \textbf{14}, R415 (2002).

\end{references}

\end{document}